# A new extremely ultrathin metasurface energy harvester and its simple modelling based on resonant half-wave dipole antenna

Alireza Ghaneizadeh[#1], Mojtaba Joodaki[*+2], Josef Börcsök[+3], Khalil Mafinezhad[#4]

[#]Department of Electrical Engineering, Sadjad University of Technology, Mashhad, 9188148848, Iran.
[*]Department of Electrical Engineering, Ferdowsi University of Mashhad, Mashhad, 9177948974, Iran.
[+]FG Rechnerarchitektur u. Systemprogrammierung, Universität Kassel, 34121 Kassel, Germany.
[1]a.ghaneeizadeh@gmail.com, [2]joodaki@um.ac.ir, [3]j.boercsoek@uni-kassel.de, [4]khmafinezhad@gmail.com

*Abstract* — In this paper we propose a novel design approach for an ultrathin metasurface energy harvester based on a surrogate model of dipole antenna. A significant advantage of this idea is the reduction of the time of the design process retrieved from a surrogate model of the resonant half-wave dipole antenna embedded in a medium. However, the design of a new electromagnetic energy harvester with a deep-subwavelength thickness ($\sim 0.004\lambda$) is the major concern of this work. The proposed structure shows an enhanced level of absorption. Decreasing the thickness of metasurface as a new classification of energy harvester, we have managed to demonstrate the stability of the efficiency. In addition, this metasurface energy harvester proved to maintain a relatively constant performance (efficiency and HPBW more than 85% and 7%, respectively) as the angle of the EM incident wave was changed over a range of 75 degrees in the transverse magnetic (TM) polarization.

## I. INTRODUCTION

Basically, it is known that natural materials have many features coming from their atom constituent and their mutual interactions, such as permeability (μ), permittivity (ε), chirality, conductivity, elasticity, viscosity, heat capacity and so on [1]. Among them μ and ε actually determine the material's electromagnetic (EM) characteristics [2]. From Lorentz force, the electric and magnetic field fluctuations lead to oscillations of charge up and downward and in the loop [2]. In fact, EM characteristics of each material are only related to their individual induced magnetic or electric dipole moments and the mutual coupling between them [3]-[5]. In such materials, due to the boundary condition effects and electric/magnetic mutual coupling in the constituent elements, analytical calculation of the susceptibility full-tensors might be a challenging task [3]-[6]. These tensors provide the most general responses of the bi-anisotropic materials to the incident EM wave, accurately [4], [5]. Furthermore, the computational cost and slowness of speed in full-wave numerical simulations may have incurred a tedious design-process of the EM structures, e.g. the antennas [7]. More recently, researchers are attempting to use simple models in order to overcome the aforementioned challenges [4], [7]-[13]. Selvanayagam and Eleftheriades came up with a circuit model for the Huygens surface [8]. Smith *et al.* have simply presented a dipole language based on the conceptual understanding of waveguide-fed metasurface antenna [9] while Liu *et al.* have presented a number of other approaches [10].

In meta-material structures, dipole responses can be created with a host medium embedded by small inclusions or meta-atoms [3]-[6]. Unlike the photonic-crystal units, which have the lattice constant about one incident EM wavelength, the meta-atoms have a periodicity being significantly smaller than this length [3]-[6]. As the size of meta-atoms are too small, they can have a macroscopic homogenized behaviour locally, and can be modelled by a simple scatter or radiation point [9], [10]. These simplest models may be helpful even when the structure becomes more complicated, but the mutual coupling tensors are required to be carefully analysed [10]. Also, meta-structures show an effective μ and ε, different from μ and ε of their constituent materials at the desired frequency [11]. However, the resonant nature of the meta-material in most of the related works is considered to be narrowband and lossy. Furthermore, the complexity and bulkiness of meta-material constructions lead to using metasurfaces instead of meta-materials [4], [6], [11]-[13].

The metasurface paradigm considers a two-dimensional meta-material act as factitious EM sheets [6], [11], [13]. These surfaces can have an abrupt field discontinuity over either side of electrically thin surface based on a generalized Snell's law [14], which can be used in engineering the EM wave-front [4]. These are assembled by such subwavelength particle textured e.g. metallic, dielectric or acoustic polarizable elements [4], [6]. The simplification and parametrization in the design process of metasurfaces can be applied in terms of the generalized sheet transition conditions (GSTCs) [4], [6], [15]. These characterizations can be accomplished by using the surface admittance (or impedance) [11], susceptibility or polarizability of metasurfaces [6], [15].

Recently, metasurface energy harvesters (MEHs) have appeared as a new collector architecture for converting the propagating wave-front to a guided wave enhancing it to reach the desired loads [6], [16]-[18]. Therefore, an MEH could be used for many functionalities such as metasurface EM energy harvesters, absorbers, data transceivers and sensors. Ideally, an advantage of deploying the metasurface in conventional antennas (excepted in array-antenna) is the equality of the physical and effective areas of the aperture [6], [16]-[18].

Over the past few years, many different scenarios have been inspected developing a maximized MEH performance e.g. multi-band, wide-incident-angle and circular polarization reception as well as polarization insensitive property [6], [17]. However, decreasing the thickness of MEH causes many problems reviewed in our work [6]. For the first time, we designed and experimentally demonstrated that EM energy harvesting can be obtained with a flexible ultra-thin (~0.004λ) non-magnetic substrate metasurface [6]. In this work, a new extremely ultra-thin MEH has been proposed. Another advantage of our idea which can be mentioned is the fact that it can decrease the design process time retrieved from the surrogate model realized with the square patches over the grounded dielectric substrate.

## II. DESIGN METHOD OF THE ULTRA-THIN METASURFACE ENERGY HARVESTER

Firstly, it should be noted that the defined frequency ranges are crowded by different communication systems such as LTE, 3G, WiFi, GSM which are of concern when it comes to EM energy harvesting, and accordingly WiFi (5.85GHz) is chosen for this research [6], [17], [18]. The schematic configuration of the proposed periodic unit-cell (with different scales) is demonstrated in Fig. 1, which is composed of a grounded substrate and a square metallic patch. The distance between these patches are (150um) in the order of 0.002 of free-space wavelength (λ). Also, two metallic vias are embedded inside the substrate which can channel the induced current to the desired loads (red areas), which mimics the input impedances of the rectifier circuit [17]. Rogers RO3010 PCB ($\varepsilon_r$=11.2 & tan(δ)=0.0022) and copper layers are used as substrate and metallic layers, respectively. Also, this MEH does not need any additional matching networks to be connected to 50 ohms loads. The thickness of the substrate and metallic layers are 0.254 *mm* and 0.5 *oz* (17*um*), respectively. Note that the total power of two loads are accumulated [17] and calculated by finite element method simulations (ANSYS HFSS).

Time-consuming full-wave numerical simulations are known to be a popular approach to the design of EM meta-structures. These simulations have been performed over the entire meta-structures to compute the scattering (S-) parameters, current or field distributions and so on. However, the EM responses of such structures are considerably affected by the location of feeds, thickness of substrate and the mutual coupling between constituent inclusions of the metasurface [6], [7].

To reduce the computational cost and speed up the process of this simulation-driven design of MEH, we have proposed a much simpler model derived from a physics-based half-wave dipole antenna which is embedded in medium with a specific permittivity. Using this simplification, the resonance frequency of the proposed MEH can be easily computed. Typically, the array antennas have a half-wavelength spacing between each element; by contrast, the distance between patches of MEH is a deep sub-wavelength (~0.002λ) range [9], [18]. Nevertheless, it should be mentioned that this simplified model cannot be used for very precise analysis.

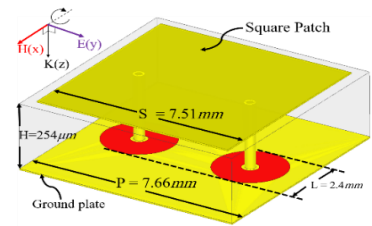

Fig. 1. The schematic configuration of the proposed periodic unit-cell.

Indeed, when the EM plane wave occurs on the polarizable particle of MS, it leads to a rise in the discontinuity of the EM fields [2], [13]. From the equivalence principles, the field distributions on either side of surface are related to the current distributions [12], [15]. Furthermore, in [19], was shown that a perfect absorption needs a non-zero physical thickness because of the existence of both equivalent sheets of the electric currents and fictitious magnetic currents [12], [20], [21]. Therefore, a perfect absorber shows both electric and magnetic responses [21]. As a result, in order to vanish the reflection and transmission completely, they need to balance the re-radiation from both of the equivalent sheets of the magnetic and electric surface currents simultaneously known as Huygens' surface [12], [13], [19]-[21].

It needs to be mentioned that, in this paper, the behaviours of both of the fields have caused forcing the AC currents towards the desired loads. Due to the grounded-substrate and impedance matching at one side of MEH, the power transmission and reflection coefficients are near zero at the resonance frequency [12], [17]. According to the result of the previous discussion, actually, we should have employed a resonant half-wave dipole antenna for design of periodicity length of the proposed MEH.

A fundamental physic rationale is used for calculating the resonance frequency (*f*) of half-wave dipole antenna [2], which is given as:

$$\frac{\lambda_m}{2} = \frac{c}{2f\sqrt{\varepsilon_r}} \qquad (1)$$

where c is the velocity of light in air and $\lambda_m$ is the wavelength in a medium with desire permittivity ($\varepsilon_r$) used in the substrate of MEH. So, the length of periodicity of our unit-cell in the resonance frequency of 5.85GHz has been 7.66mm, from (1).

## III. FULL-WAVE SIMULATION RESULTS

To better illustrate the physical conception of the energy harvesting, the simulated electric field and surface current distributions of the unit-cell are plotted in the Fig. (2) and (3) at the resonance frequency, respectively. It is perceived that the surface currents and electric fields are mostly distributed around the metallic vais and lossy environments between the neighbouring patches, respectively. This effect is due to the large coupling, which exists between the neighbouring patches [6] with a distance of roughly 0.002λ (near-field region of the cells).

Accordingly, the level of absorption is defined as A (f) =1- R(f) - T(f). R (f) and T (f), represent the power reflection and transmission coefficients, respectively [17], [20], [21].

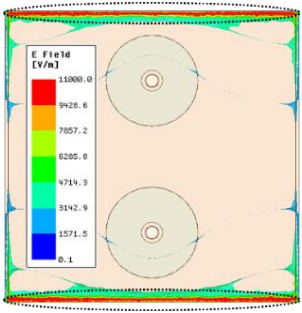

Fig. 2. Full-wave simulation result of the electric field distribution for the periodic unit-cell of the proposed MEH at the resonance frequency.

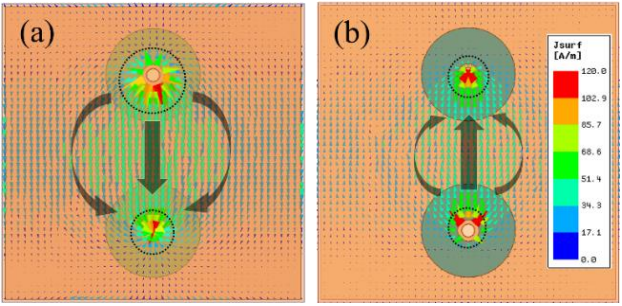

Fig. 3. Full-wave simulation result of the surface current distribution: (a) below the patch layer; (b) above the ground plate at the resonance frequency.

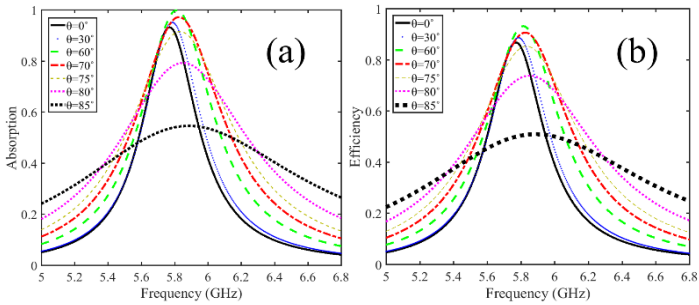

Fig. 4. Full-wave simulation results of the: (a) Absorption; (b) Efficiency of the proposed structure.

Finally, to analyse the angular stability of the proposed structure, the numerical results of the absorption and efficiency of MEH are inspected under a different oblique incident angle, as shown in Fig. 4. As it is observed, the maximum level of efficiency is above 92% of the accessible received power to the MEH at θ = 60°, as a result, a suitable impedance matching is achieved not only at the interface between the free space and top surface of MEH, but also between the 50 ohms loads and bottom surface of MEH. It should be noted that in this research the incident plane-wave is TM polarized with respect to MEH surface.

## IV. CONCLUSION

In summary, in this paper, a novel approach to design of an extremely ultrathin MEH has been thoroughly explained. From the results, it can be found that the proposed surrogate modelling of MEH to approximately design resonance frequency has not only been concurred to numerical full-wave simulations, but also the proposed ultra-thin (~0.004λ) MEH with angular sustainability is a good candidate for 2D-isotropic EM energy harvesting [6]. Moreover, increasing the angle of incidence from 0 to 75 degree has caused to half power bandwidth ratio (HPBW) increase from 7% to 12%. Also, these structures have actually shown a perfect level of absorption.